# Nanomechanical strain concentration on a 2D nanobridge within a large suspended bilayer graphene for molecular mass detection

Julien Chaste[1*], Amine Missaoui[1], Amina Saadani[1], Daniel Garcia-Sanchez[2], Debora Pierucci[1], Zeineb Ben Aziza[1], Abdelkarim Ouerghi[1]

1) Centre de Nanosciences et de Nanotechnologies, CNRS, Univ. Paris-Sud, Universite Paris-Saclay, C2N – Marcoussis

2) Sorbonne Universités, UPMC Univ. Paris 06, CNRS-UMR 7588, Institut des NanoSciences de Paris, F-75005, Paris, France

* julien.chaste@c2n.upsaclay.fr

The recent emergence of strain gradient engineering directly affects the nanomechanics, optoelectronics and thermal transport fields in 2D materials. More specifically, large suspended graphene under very high stress represents the quintessence for nanomechanical mass detection through unique molecular reactions. Different techniques have been used to induce strain in 2D materials, for instance by applying tip indentation, pressure or substrate bending on a graphene membrane. Nevertheless, an efficient way to control the strain of a structure is to engineer the system geometry as shown in everyday life in architecture and acoustics. Similarly, we studied the concentration of strain in artificial nanoconstrictions (~100 nm) in a suspended epitaxial bilayer graphene membrane with different geometries and lengths ranging from 10 to 40 µm. We carefully isolated the strain signature from µ-Raman measurements and extracted information on a scale below the laser spot size by analyzing the broadened shape of our Raman peaks, up to 100 cm$^{-1}$. We potentially measured a strong strain concentration in a nanoconstriction up to 5%, which is 20 times larger than the native epitaxial graphene strain. Moreover, with a bilayer graphene, our configuration naturally enhanced the native asymmetric strain between the upper and lower graphene layers. In contrast to previous results, we can achieve any kind of complex strain tensor in graphene thanks to our structural approach. This method completes the previous strain-induced techniques and opens up new perspectives for bilayer graphene and 2D heterostructures based devices.

**Introduction**

The rise of 2D materials and atomically suspended thin materials have brought new insights into mechanical systems. Graphene is among the strongest materials due to its very strong in-plane C-C bond between atoms, which results in high strain ratios in the linear elastic range and Young's modulus higher than one terapascal. On the other hand, graphene is the thinnest material available with a high geometrical aspect ratio, which leads to both a low mass and low spring constant [1,2]. Because of these abnormal peculiarities, almost all the intrinsic properties of suspended 2D membranes are highly modified by an applied strain.

Recently, various experiments in nanomechanics, optoelectronics or thermal transport have been dedicated to the efficient tunability of 2D material properties when submitted to an applied stress[3–5]. Subsequently, it is interesting to understand the processes in action during a unique chemical reaction by considering the mass detection of molecules from few Daltons to hundreds of kDaltons. The mass spectrometry within this range is challenging and nanomechanical resonators are the viable solution. While a carbon nanotube can reach the 1 Daltons atom detection ($10^{-24}$ grams) [6], it is not suitable for broad applications due to its one-dimensionality, its bad integration at large scale and the low-temperature experiments. Contrarily, the 2D epitaxial graphene is compatible with scalable fabrication and offers the possibility to create ultralight resonators with the large surface at ambient temperature. However, to improve the mass detection of molecules and ad-atoms on these resonators, it is necessary to increase its internal strain, from typically less than 0.05% to up to a few percents.

The static strain is similar to a pseudo-scalar potential that modifies the band structure of graphene or other 2D membranes. The stress is used to tune the 2D semiconducting material band gap and to open an energy gap in the semimetal bilayer graphene.[7–13] This electronic tunability has led to applications in photovoltaics [13] and optical quantum emissions.[14,15] Strain tensor engineering represents also the basis for other concepts in atomically thin materials, such as artificial stacking of 2D materials. In these thin heterostructures, the mechanics entirely govern the layer interaction for vertical heterojunctions[16–18] or lateral junctions[19,20]. In addition, the concentration or release of strain in nanomechanical graphene resonators by simple nanostructuring boosts the development of the novel mass/force detection [1,6,21] and also the non-linear mechanical devices. Therefore, nanostructuring-strain duos play an important role in the dissipation and dilution of mechanical resonators.[22] This emphasizes the importance of strain engineering at the nanoscale (<10 nm) to tailor the geometry of a suspended mechanical resonator.

The simplest and most versatile tool to measure strain in suspended 2D materials is Raman spectroscopy[23,24]. Measurements with scanning tunneling spectroscopy[13,25] or transmission electron microscopy[19] are spatially resolved but difficult to implement for suspended 2D materials on a chip and are limited experimentally and environmentally. Because of the finite laser spot size, strain measurement in nanostructures with Raman spectroscopy has important limitations, though recent efforts have attempted to resolve strain details at the nanoscale [26,27]. Specific methods have been used to separate and discriminate the doping and strain components measured on a substrate [23] and on suspended structures [28]. Previous studies only considered the simple Lorentzian Raman peak shape; here we show how to extract strain information from more complex peak shapes at the nanoscale. Graphene Raman peaks have been investigated over suspended bilayer epitaxial graphene in order to measure strain at the nanoscale. We created specific nanostructures where the strain could be concentrated on a nanoconstriction, using a relatively fast and simple method. Strain measurements were compared with finite element mechanical analyses. We possibly measured a strain concentration of 5 % at a nanoconstriction with strong asymmetry shown between the top and the bottom graphene layers. We have demonstrated the potential of this method to obtain any kind of strain pattern.

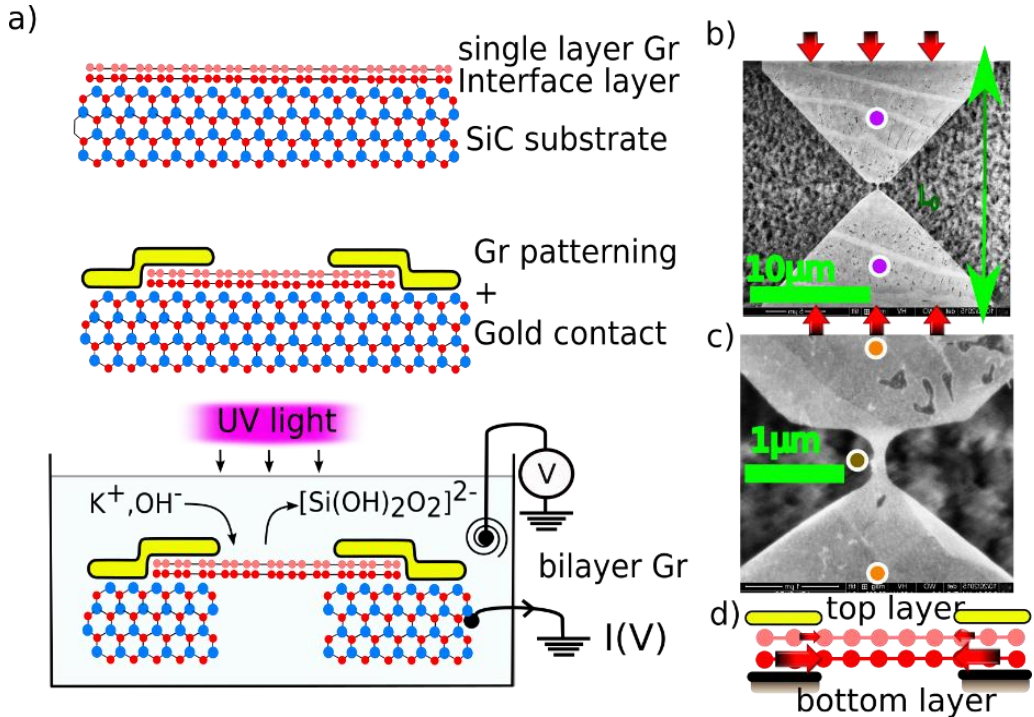

**Figure 1**: **A graphene 2D anvil cell a) Sample fabrication process from top to bottom:** growth of an epitaxial bilayer graphene on the SiC substrate, graphene and metallic pads e-beam nanopatterning and the final photo-electrochemical etching under UV light to release the graphene from the substrate **b) b)** SEM images of suspended bilayer epitaxial graphene structures with a 2D diamond anvil cell geometry within a membrane of side $L_0$= 20 μm **c)** zoom on the nanobridge in the middle of the membrane (W=160 nm, L=600 nm). **d)** A diagram of the doubly clamped suspended epitaxial graphene bilayer initially under compression.

## Results

For sample manufacture, we used epitaxial bilayer graphene grown by sublimation on a SiC substrate at 900°C as in Figure 1a. We previously demonstrated our experience in nanostructuring and suspending a complex design in our graphene over a large distance [24]. By e-beam lithography, we patterned the graphene into large structures composed of two opposed triangles connected by a small nano-bridge. We clamped the graphene on both sides with long bars of gold, which were, at the same time, the electrical contact for the photoetching, the graphene clamps, and the etching mask. In this way, we etched a few microns of SiC underneath the graphene using a photoelectrochemical method [29]. We obtained a very large membrane of suspended bilayer graphene, which is a 2D equivalent of the diamond anvil cell. In Figures 1b and 1c, we present a nanoconstriction of width 160 nm (W), and length 600 nm (L), right in the middle of a bilayer graphene membrane with a total length of 20μm ($L_0$). If we consider the native strain along this structure, it is important to note that this actually works as a real anvil cell. The initial stress in the graphene is the native biaxial compressive stress $T_0$ induced by the epitaxial growth: -2.27 GPa (strain $\varepsilon_0$ = -0.2%), mainly on the bottom layer[30]. In our material, the strain was asymmetric in the vertical direction: the upper layer was almost released mechanically. During device suspension, the native strain of the whole membrane was redistributed around the central nanoconstriction. This induced a high strain concentration in this region, which increased the vertical strain asymmetry.

Many structures were created in parallel (around 100) to achieve a sufficient number of samples with a success rate of 10% or below. Two steps were quite critical, the photoelectrical etching itself and the subsequent drying. We took care to avoid bubbles during the etching in KOH at 50°C and under UV light >0.5mW. Details are described elsewhere [24]. We etched around 8 μm of SiC with this technique at a speed of 2μm/hour. The sample was then rinsed in water a few times and dried with a critical point dryer. Only a few samples with $L_0$=40μm were achieved with a lifetime of around one week. Samples of length $L_0$=20μm or below were much more stable and easily made.

**Strain signature in Raman.**

We then focused on the μ-Raman measurements and the strain concentration measurements around these nano-constrictions. We used the Raman signature and the 2D peak shape to distinguish bi- and tri-layer graphene areas (contrasted lines in the e-beam image, see Figure 1b), intrinsic to the sublimation method[31]. A typical Raman spectrum is shown in Figure 2a with peaks D, G, 2D, and 2D'. Raman is an efficient tool for strain measurement if a specific methodology is used to distinguish the Raman spectrum peak evolution under strain, doping, temperature or layer number. Stressed graphene has a very specific Raman signature; the peaks have a relative position $\Delta\omega_G=\omega_G.(\gamma.\varepsilon_H \pm 0.5.\beta.\varepsilon_A)$, where $\varepsilon_H = \varepsilon_{11} + \varepsilon_{22}$ is the hydrostatic strain, which represents the sum of the elongation in the two main orthogonal directions 1 and 2 and $\varepsilon_A = |\varepsilon_{11} - \varepsilon_{22}|$. The Gruneïsen parameter, $\gamma$, is 1.8 for the G peak[32] and the shear deformation potential of graphene, $\beta$, is ~0.8-1. When the strain tensor is anisotropic in the graphene plan ($\varepsilon_A \neq 0$) the Raman peaks shift and split into two peaks G+ and G- [32–34].

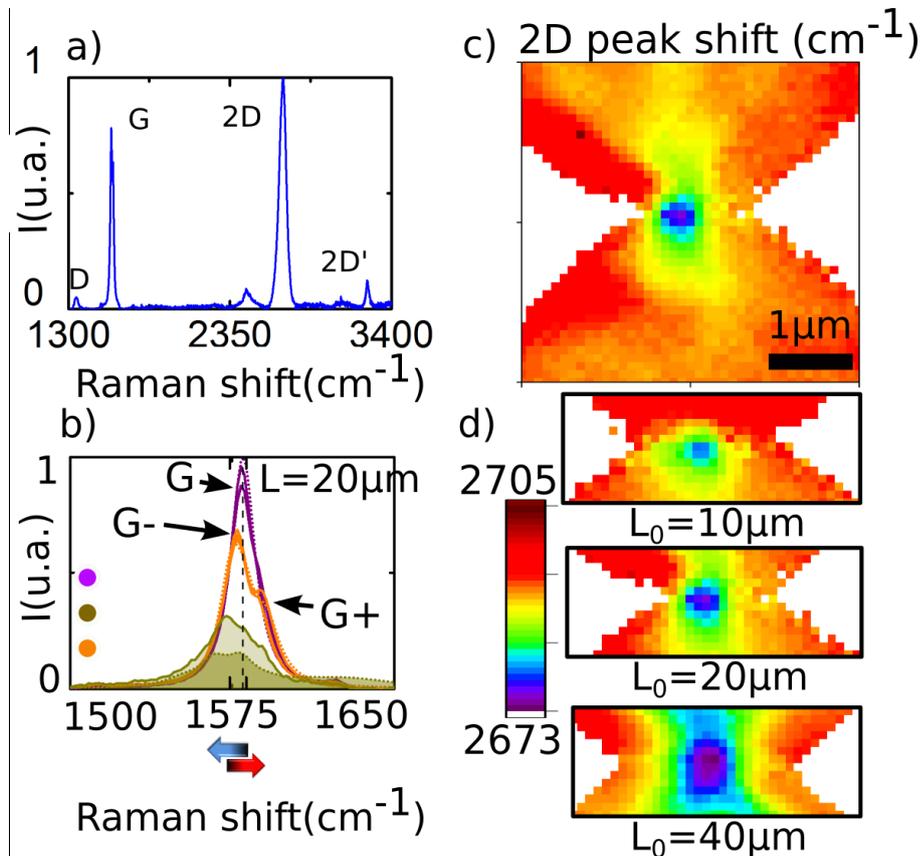

**Figure 2**: **A strong Raman singularity at the nanoconstriction a)** A typical Raman spectrum **b)** G peak measurements at 5 different points on the 20μm membrane with respect to centrosymmetry. Corresponding colors are presented in Figures 1a and 1c and simulations are in dashed lines of the same color. By approaching the structure center, the curve shape progressively transforms from a Voigt curve to a broadened and complex shape with contributions above and below 1580cm$^{-1}$(vertical black dash line). **c)** Raman mapping around the nanoconstriction, in Figure 1, for the 2D peak position after a simple Voigt peak fitting. A strong shift of the peak position is observed around the nanoconstriction. A more rigorous 4 Lorentzian fit gave a similar but noisier result (see SI). **d)** Different mappings over structures with different $L_0$. The central redshift increased systematically with $L_0$

In Figures 2, we present the Raman mapping of the 2D peak position, around the nanoconstriction, for 3 different membranes of lengths $L_0$, varying from 10 to 40µm and at room temperature. We observe a global peak shift around the nanoconstriction. This shift increases with $L_0$. Similar features were observed in all our samples. In Figure 2b, we present specific measurements of the G peak shape at five places along the membrane for $L_0 = 20$µm (these five spots are centrosymmetric to the small bridge and shown by colored circles in Figures 1b and 1c). Far away from the nanobridge, in purple, the peak is similar to a Lorentzian curve at ~1580 cm$^{-1}$. It defines a no-strain references point for our intensity, width and position peak Raman simulations, and is quasi-similar to what we can obtain in a membrane without nanoconstrictions. Closer to the nanoconstriction (orange points), the peak seems to split into two, G$^+$ and G$^-$ peaks on each side of the 1580 cm$^{-1}$ position. At the central point (brown point), the G peak shape is quite complex and broadens from around 1530 cm$^{-1}$ to 1630 cm$^{-1}$. From the $L_0$ dependence, the peak shape and other peculiarities of our signal, we have isolated Raman strain signature around our nanobridge without ambiguity;

    1) In Figure 2d, the Raman peak shift increases with the sample geometry $L_0$. Among the strain, doping or heating contributions, only the strain is correlated with the length of such a structure as shown by our strain simulation. Outside a mesoscopic regime, a strong doping variation is not correlated with geometry, especially $L_0$, but related to the material itself. Only a negligible edge doping contributions can occurs[28]. For heat, in this double triangular shape, the thermal resistance far away from the center is negligible due to the large width and due to the thermal heat transfer to air [35]. It results in the main thermal resistance contribution to be limited to a region of few micrometers around the nanoconstriction and to be quasi-independent of $L_0$.

    2) The laser power calibration shows a laser heating effect at the central point only for power above 100µW. In consequence, we have proceeded the measurements below with the laser power of 50µW avoiding heating issues, as shown in Figure 3b.

    3) An efficient way to distinguish the effect of strain, temperature and doping on the Raman spectrum is to determine the ratio between the 2D peak shift and the G peak shift. For example: $\Delta\omega_{2D}/\Delta\omega_G$ is 0.7, at its maximum, in the case of doping (this can even be negative for n-type doping) [36], 1.7 for a thermal variation (~ $\omega_{2D}/\omega_G$) [37] and over 2 for strain (due to an additional contribution from Gruneïsen parameters) [32–34]. To measure this ratio, we plotted the Lee et al. diagram[23] in Figure 3a around the nanoconstriction for $L_0= 40$µm. The slope of 2.2 indicates a strain variation. In order to take into account the non-Lorentzian nature of our spectrum, we used the weighted position of the peaks. Our simulation reproduces the slope and these atypical shapes for both the G and 2D peaks. The procedure is described later.

    4) Splitting of the G peak is not exclusively related to the strain but only a planar anisotropic strain modifies the hexagonal shape symmetry of the graphene mesh, and the Raman response that ensues depends on the axis of polarization, in contrast to other scenarios (strongly asymmetric vertical doping, the presence of an even number of graphene layers [38–41], an additional molecular interaction [42] and a confusion with the D' peak at 1620 cm$^{-1}$). In Figure 3c, for a membrane of 10 µm, we analyzed the G peak splitting, at a point away from the

nanobridge, and represented the intensity area ratio between G+ and G- as a function of the polarization angle. We see a modulation of the ratio between 1 and 2. This value is compatible with a two-layer epitaxial graphene where strain is dominant in only one layer [30]. Because our data are usually taken without any polarization, unless it is notified, it is not possible to define the crystallinity orientation (zig-zag, armchair) on the nanobridge itself.

5) In Figure 3d we show the Raman 2D' peak shift (standard position 3250 cm$^{-1}$) along with an "$L_0$ = 10 μm" membrane. This peak has been considered to be useful for strain determination [43] because it is less subject to doping or temperature dependence. We obtained the expected peak shift ratio ($\omega_{2D'}/\omega_{2D}$) and peak width ratio from graphene under strain[43].

6) In Figures 2b and 4b, we measured large peaks, at the nanobridge position, with spectral contributions above and below the medium position, ~1580cm$^{-1}$. Temperature and doping variation alone hardly explain this, considering that it is difficult to have the coexistence of cold and hot regions or of a naturally strong doping of opposite sign within a distance of a few hundred nanometers. This point suggests that the mechanics within our graphene are potentially more complex, as shown by our simulation.

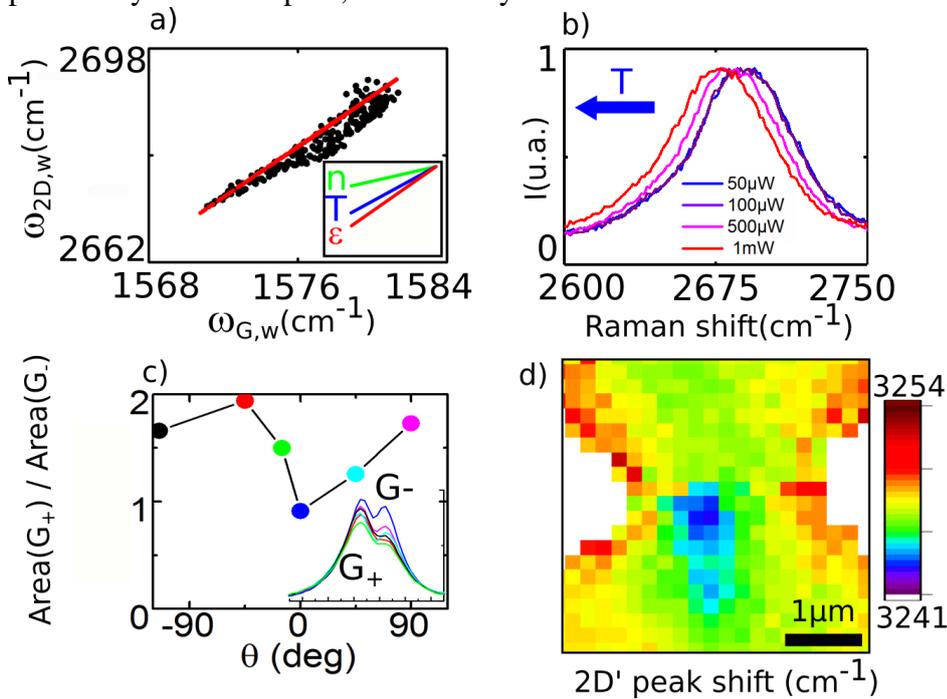

**Figure 3: The strain signature a)** the relation of the 2D weighted peak position to the G peak for the 40μm membrane around the constriction. The red points are the simulated stresses for bilayer membranes and the green, blue and red slopes, in the inset, show, respectively, slopes of 0.7 (doping),1.7(heating) and 2.2(strain). **b)** 2D peak shape function of the laser power. A small heating effect is observed for higher powers than used in our measurements. **c)** Polarization dependence of the G peak shape when two peaks are clearly visible. We observed an area ratio dependence as expected by strain measurements. In the inset, we present the G peak at the different angles with the corresponding colors. **d)** Another strain signature with 2D' peak mapping for the 10μm membrane and its position shift around the nanoconstriction.

**Strain signature in the peak shape with an asymmetric bilayer graphene model.**

We have demonstrated the dominant strain variation around the nanoconstriction, and our ability to extract strain properties from our Raman measurements. We will now focus on one important behavior, far away from the nanobridge, in a quasi no-stress region, where we measured typical Lorentzian curves but whose features, at the constriction position, were quite complex. This was related to the high strain gradient along the nanobridge over a distance below the diffraction limit of our optical measurement method. This means that important information about strain is to be found in the Raman peak shape itself. In order to recover part of this information from our data, we simulated the mechanical behavior of our membrane with different initial strains and geometries and with clamped edges. The local strain distribution and the resulting local spectral Raman response were spectrally convoluted with the Gaussian distribution of the collected photons. Without strong local variations, this convolution procedure rebuilt the original Lorentzian (far from the bridge). At a local high gradient of frequency shift or even at the graphene edge, the result was no longer necessarily a Lorentzian curve or a normalized area but it could be a very complex feature as measured. We extracted the expected peak shape at each point in order to obtain a good fit with our data and to quantify the strain in our system in the nanometer range. As a result, with a small deviation from data, we were able to fit the five spectra in Figure 2b within our asymmetrically strained bilayer graphene membrane and to extract the different strain contributions. Similar studies were done for the G and 2D peak shapes of a 40µm membrane in S3. Finite element analyses were done using Comsol and confirmed in parallel with Ansys software. The Gaussian spot size was measured to be 450 ±50 nm, calibrated *in situ* at the graphene edge for the $L_0$ = 20 µm membrane measurement, (see S2), and we also determined it to be around 500 nm directly through the simulations, as a free parameter. Until now, we had intentionally avoided mentioning the two-layer contributions to our Raman peak shape.

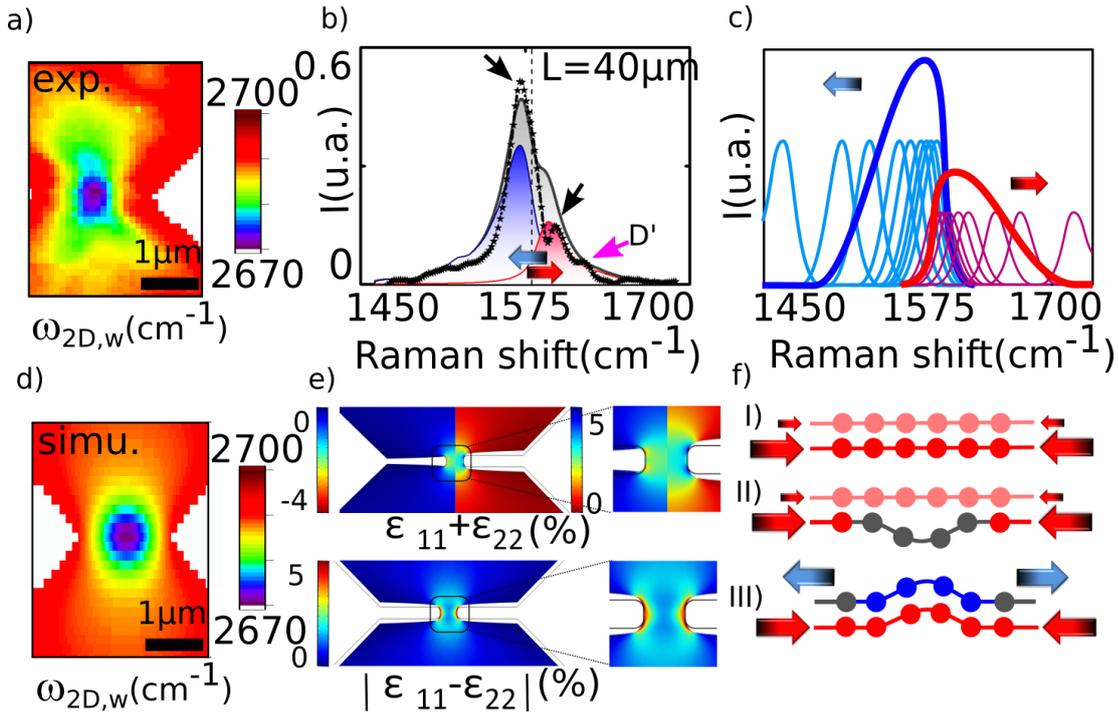

**Figure 4: Strain simulations and fitting on a 40µm membrane a) and d)** Experimental and simulation of the weighted 2D peak position (as in Figure 3a) along the nanobridge constriction for the 40µm membrane **b)** Raman measurement at the center of the nanoconstriction for a 40µm membrane with a maximal strain concentration (line with stars) and simulation in black line, with the stressed layer contribution in blue and the compressed layer in red. **c)** Diagram explaining the Raman complex shape by an accumulation and convolution with the Raman spot of all the local Lorentzian peaks with different local strains. In blue and red, we present the stressed and compressed contributions, respectively, from each different layer. **e)** Comsol simulation of $\varepsilon_H$ of each layer (with $\varepsilon_0$ of -0.06% on the right and 0.08 % on the left) and $\varepsilon_A$ for $\varepsilon_0 = 0.08$ % corresponding to a) and around the constriction of the 40μm membrane, which defines the local strain and Raman response. **f)** In an asymmetric bilayer situation, where a single layer is compressed, only a scenario with some buckling and where the two layers are curved together can explain both our stressed and compressed measurements (scenario III)

To improve the strain concentration, in Figure 4b, we plotted the G peak shape at the center of a membrane with a large support of $L_0=40$ μm. We observed an even more pronounced broadening of the G peak, up to $100 cm^{-1}$, as compared to shorter membranes and a separation into two Raman peaks. We naturally supposed the spectral contribution below $1580 cm^{-1}$ to be due to positive strain (blue) and above $1580 cm^{-1}$ to be related to a compression (red), as represented in Figure 4c. This asymmetrical strain is the induced image of the native strain asymmetry between the two layers of our epitaxial bilayer graphene. It is impossible to report our results with only the simulation of a single monolayer. In fact, in this singular system, the natural strain and stress distributions are applied homogeneously by the SiC substrate on the bottom layer and the upper layer which is relatively unstressed, or at maxima, negligibly compressed [30]. Initial stresses are released during etching and create a relative movement of the layers. In the first step of our analyses, we considered two independent layers of graphene in the elastic regime and without any buckling stress. A lateral displacement - sliding of the individual graphene layers, is expected due to the weak Van der Waals forces friction or adhesion between the two layers[44], which is even weaker in an incommensurate rotation state [45]. It has been previously shown that a bilayer graphene under lateral strain can lose his inversion symmetry breaking and the relative stacking of the two layers, inducing a two independent layer system [46]. Eventually, it has also been shown in the case of multiple layers graphene [47,48].

The Young 's modulus of graphene is around 1Tpa, with strain dependence defined in ref [49] and layer thickness is 0.33nm. For our data to fit qualitatively well for a membrane with $L_0=$ 40μm, we used an initial strain $\varepsilon_0$ of -0.06% and +0.08 % for each layer and simply added their spectra matrix (Figure 4d and 4e). In parallel, this result was also obtained by well-fitting the G and 2D weighted peak positions, as in Figure 3a (and the maximum peak position, as in the SI, to determine β). We plotted the resulting $\varepsilon_H$ and $\varepsilon_A$, in Figure 4e, for the top and bottom layers. $\varepsilon_H$ and $\varepsilon_A$ are 2.3% and 1.4%, respectively, in the middle of the nanobridge for the stressed layer. This is around the mean strain applied on the bridge and corresponds to $\varepsilon_{11}=1.9\%$ > $\varepsilon_{22}=0.4\%$. From these simulations, we achieved strain concentration at the nanoconstriction with a gain at least >20 in the center of our constriction and a clear transition from biaxial to uniaxial strain.

It is noticeable that closer to the edges, we have maxima at $\varepsilon_A=5.6\%$> $\varepsilon_H=5.0\%$, which means $\varepsilon_{11}=5.3\%$ > $\varepsilon_{22}=$ - 0.3%; the orthogonal strain applied at this point is of the opposite sign. This state is difficult to obtain experimentally with other methods. We have a high strain gradient and a strong strain asymmetry, with a difference of 5 to 9% between the two layers. Similarly, we have also simulated the $L_0=20$μm membrane with fitted value $\varepsilon_0$ of +0.02% and -0.12 %. We noticed, in each sample, negative and positive contributions, which were close to but never exceeded the |-0.2%| strain in the bottom layer of native epitaxial graphene, even by adding their module. This means we have obtained a realistic initial strain value. This is an unprecedented achievement and demonstrates the powerful ability of nanostructuring to control strain in suspended graphene, even in an asymmetric configuration.

## Discussion
Until now, we have analyzed our Raman spectrum data of a bilayer graphene as the addition of two monolayers, mechanically independent. Now, we complete our model with the possible bending of the membrane under high strain and explain why we have systematically one layer under a low extension at the initial stage ($\varepsilon_0 > 0$). Since the bottom layer is clearly under compression, we attribute this positive strain to the upper layer. This can be surprising since it is not necessarily the case; the upper layer's initial strain must be unstressed or tend to be negative in the case of small interactions with the compressed lower layer. We have to consider the interactions between layers, especially the effect of buckling strain on both layers. With a graphene monolayer, if enough compressive strain is applied to the layer, it will generally induce buckling and a strong reduction in the Raman strain signature, as in [12]. A symmetrically-stressed graphene bilayer improves the rigidity of the system, but will not necessarily avoid this bending. Lowering the system energy minima to equilibrium will eventually lead to 3 scenarios, presented in Figure 4f: i) No buckling; as previously mentioned, if each membrane stays mechanically independent, resulting from only an apparent compression and no extension term. ii) Buckling of one layer only; for this second scenario, the lower layer bends downward under compression. With small interlayer sticking, the physical layer separation increased [50]. This was not our case since it implied a complete release of the stress-energy into bending and a residual Raman strain shift. iii) Buckling of the 2 layers; finally, with enough sticking between layers and no physical separation, the lateral support offered by the second layer reduces the buckling strain. A bending of the lower layer inevitably induces a similar curvature on the upper layer but with opposite stress. This implies an equilibrium between compression, bending, and stress. The strain around the nanobridge will appear with both stress and compressive layer signatures, as measured and it explains our observations quite well. This graphene extension is not related to a random sliding of the gold clamp because this effect increases with length $L_0$ and appears in all of our samples.

By approximating our system to two graphene monolayers, we assume doping induces each layer to be shifted in energy and strain induces each layer to shift in k space. We put aside the fact that Raman behaviors are more complex when discussing strain and doping effects on the band structure of bilayer graphene in a vertically asymmetric configuration. Nevertheless, in ref [8], it can be seen that the influences of asymmetrical doping or strain on the band structure of a bilayer are quite similar, to a first approximation, to their influence on two independent graphene layers. In our model, we neglected the appearance of any plasmonic resonance, especially at the nanobridge, and consequently their coupling to the Raman signal. We demonstrate the absence of any heating as plasmons heating. We neglect the possible Purcell effect or any related electrostatic effect due to plasmons. We also avoid the possibility of optical coupling with a metallic nanoparticle randomly deposited at the nanobridge position because it is not in accordance with a strain proportional to $L_0$ without any negative occurrence in our measurements.

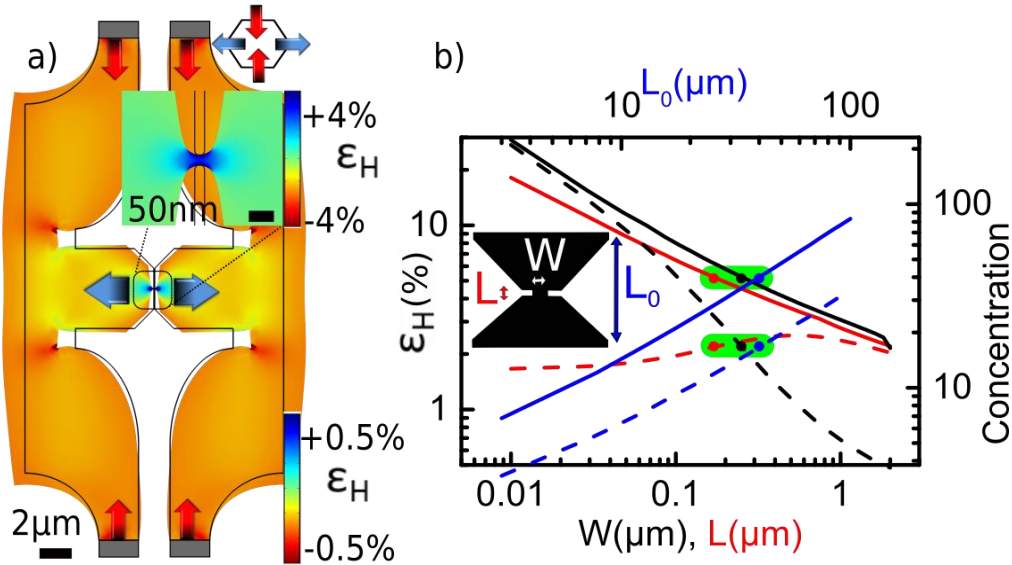

**Figure 5: Versatility of nanoengineering to create multiple strain configurations. a)** Simulation of a single layer graphene membrane with a hexagonal geometry. This results in a possibility to invert the sign of the strain or to apply rotation geometry in the nanobridge using the same initial clamped conditions (here a compression). By comparison with the diamond anvil cell with similar clamping length, the efficiency is reduced only 2-fold. The global length $L_0$ is 25µm and the nanobridge is 170nm by 250nm (W and L for the $L_0$=40µm membrane). **b)** The hydrostatic strain $\varepsilon_H$ as a function of $L_0$, L and W. The geometry present in b, is set by default in a. It is possible to improve the concentration up to 50 in the nanoconstriction by simple geometrical considerations.

Using our simple approach, it is possible to achieve any kind of complex strain tensor in the 2D material. To demonstrate this, we extended our simulations to more complex geometries in the case of a layer graphene. In Figure 5a, we present this almost hexagonal shape. This geometry possesses the peculiarity to transform a compression into highly efficient orthogonal stretching. If we compare this structure to our anvil cell geometry, taken as a reference with equivalent length W, L, $L_0$, $\varepsilon_0$ and anchoring length (see SI), we obtain an orthogonal rotation of the strain on the nanobridge with an opposite sign and with 50% efficiency. Starting from the geometry in Figure 5a (reference points in the green area), we also achieve a simulation for different W, L, and $L_0$ and trace the hydrostatic strain for its value at the nanobridge center and at its maximum value along the nanobridge in Figure 5b. It is possible to achieve strains above 10% with a concentration above 50 in some cases and demonstrate the powerful ability of our approach of nanostructuring to engineer the strain in graphene nanobridges with a rotating frame, to modify the strain sign and improve the concentration up to the limit of the elastic regime of graphene.

**Conclusion:**

To sum up, we have made a very large epitaxial graphene bilayer membrane with a nanobridge constriction to concentrate the native strain of the whole structure into a small area. Hence, we successfully obtained nanoresonators with small masses, very high strain and mechanical frequency, which improve the performance of ad-mass detection at room temperature. In addition to the peculiar geometry of our devices, we were able to measure the strain of our system using complete µ-Raman spectroscopy. We found out that the main contribution to Raman wide peak broadening is due to the high strain on the nanobridge. We used a new analysis of the Raman peak shape to simulate the strain distribution along the structure. We also emphasized the possibility of concentrating the strain in suspended graphene to create and control the vertical strain asymmetry and to change the strain orientation by simple geometrical considerations. We do not consider to be limited by the native strain existing in epitaxial graphene. Moreover, it is possible to combine our technique with the previous methods of strain engineering, in order to amplify and tune this effect on the same device; for example, to tune artificially the curvature of a membrane by flexion or tip indentation[50]. It is worth to note that we integrated into our model a simplified interaction between layers considering two layers independent system. This is valid if we consider that the native strain asymmetry reduces strongly the layer interactions when we measure an asymmetric Raman peak shape under high strain. Further experiments can be done to investigate the low-frequency Raman modes (<50cm$^{-1}$) to carry out to the vibrations and interactions between layers and to conduct systematic polarized measurements[51]. Finally, for nanomechanical applications, the application of a strain of 5% on a suspended 2D nanomembrane is potentially also a rare achievement. This could surpass actual mass sensing experiments with optomechanical measurements[6]. Our results open a new approach to strain engineering in other types of 2D materials such as very large homogenous flakes of graphene obtained by chemical vapor deposition or with a semiconducting material, such as $MoS_2$ or $WSe_2$ for electrons or light manipulation.[6] Strain engineering by nanostructuring is also at the heart of recent important results on dissipation dilution in mechanical resonators[22].

## Methods
### Raman set-up
The Raman spectroscopy is done with a commercial Renishaw set-up, Measurements were made at ambient temperature and pressure. The laser spot size was calibrated in situ (see SI) at around 350nm. The platform resolution is 100nm the plane (i.e., x and y). The laser is at 532nm and the spectral resolution is around 1.3cm$^{-1}$ with a grating of 1800 lines/mm. It allows us to resolve ultimately a peak position at less than 0.1cm$^{-1}$ when the peak width is already ~ 6 to 10 cm$^{-1}$. (see ref [6]).

### Samples
Many structures were fabricated in parallel (around 100) to achieve a sufficient number of samples with a 10% of success or below. Two steps are quite critical, the photoelectrical etching itself and the afterward drying. During the etching, we took care to avoid bubbles during the etching in KOH at 50°C and under UV light >0.5mW. Details are described elsewhere[24]. We etched around 8 µm of SiC with this technic at speed of 2µm/hour. The sample was then rinsed in water a few times and dries with a critical point dryer. Only a few samples with $L_0$=40µm were achieved with a lifetime around one week. Samples with length $L_0$=20µm and below were much more stable and easily done.

**Acknowledgments:** This work was supported by ANR H2DH grants and by the French Renatech network.

## REFERENCES

(1)     Chaste, J.; Eichler, A.; Moser, J.; Ceballos, G.; Rurali, R.; Bachtold, A. A Nanomechanical Mass Sensor with Yoctogram Resolution. *Nat. Nanotechnol.* **2012**, *7* , 301–

304.

(2)     Moser, J.; Güttinger, J.; Eichler, A.; Esplandiu, M. J.; Liu, D. E.; Dykman, M. I.; Bachtold, A. Ultrasensitive Force Detection with a Nanotube Mechanical Resonator. *Nat. Nanotechnol.* **2013**, 8, 493–496.

(3)     Blees, M. K.; Barnard, A. W.; Rose, P. A.; Roberts, S. P.; McGill, K. L.; Huang, P. Y.; Ruyack, A. R.; Kevek, J. W.; Kobrin, B.; Muller, D. A.;McEuen P.L. Graphene Kirigami. *Nature* **2015**, *524* , 204–207.

(4)     Amorim, B.; Cortijo, A.; de Juan, F.; Grushin, A. G.; Guinea, F.; Gutiérrez-Rubio, A.; Ochoa, H.; Parente, V.; Roldán, R.; San-Jose, P.; Schiefele, J; Sturla, M; Vozmediano, M.A.H. Novel Effects of Strains in Graphene and Other Two Dimensional Materials. *Phys. Rep.* **2016**, *617*, 1–54.

(5)     Goldsche, M.; Sonntag, J.; Khodkov, T.; Verbiest, G. J.; Reichardt, S.; Neumann, C.; Ouaj, T.; von den Driesch, N.; Buca, D.; Stampfer, C. Tailoring Mechanically Tunable Strain Fields in Graphene. *Nano Lett.* **2018**, *18* , 1707–1713.

(6)     Chaste, J.; Missaoui, A.; Huang, S.; Henck, H.; Ben Aziza, Z.; Ferlazzo, L.; Naylor, C.; Balan, A.; Johnson, A. T. C.; Braive, R.; Ouerghi, A Intrinsic Properties of Suspended $MoS_2$ on $SiO_2$/Si Pillar Arrays for Nanomechanics and Optics. *ACS Nano* **2018**, *12* , 3235–3242.

(7)     Choi, S.-M.; Jhi, S.-H.; Son, Y.-W. Controlling Energy Gap of Bilayer Graphene by Strain. *Nano Lett.* **2010**, *10* , 3486–3489.

(8)     Crosse, J. A. Strain-Dependent Conductivity in Biased Bilayer Graphene. *Phys. Rev. B* **2014**, *90* .

(9)     Nanda, B. R. K.; Satpathy, S. Strain and Electric Field Modulation of the Electronic Structure of Bilayer Graphene. *Phys. Rev. B* **2009**, *80* .

(10)    Zhang, Y.; Tang, T.-T.; Girit, C.; Hao, Z.; Martin, M. C.; Zettl, A.; Crommie, M. F.; Shen, Y. R.; Wang, F. Direct Observation of a Widely Tunable Bandgap in Bilayer Graphene. *Nature* **2009**, *459* , 820–823.

(11)    Xu, K.; Wang, K.; Zhao, W.; Bao, W.; Liu, E.; Ren, Y.; Wang, M.; Fu, Y.; Zeng, J.; Li, Z.; Zhou, W.; Song,F.; Wang, X.; Shi, Y; Wan, X.; Fuhrer, M.; Wang, B.; Qiao, Z.;Miao, F.; Xing, D. The Positive Piezoconductive Effect in Graphene. *Nat. Commun.* **2015**, *6*, 8119.

(12)    Castellanos-Gomez, A.; Roldán, R.; Cappelluti, E.; Buscema, M.; Guinea, F.; van der Zant, H. S. J.; Steele, G. A. Local Strain Engineering in Atomically Thin MoS2. *Nano Lett.* **2013**, *13* , 5361–5366.

(13)    Li, H.; Contryman, A. W.; Qian, X.; Ardakani, S. M.; Gong, Y.; Wang, X.; Weisse, J. M.; Lee, C. H.; Zhao, J.; Ajayan, P. M.; Li, J.; Manoharan, H.C.; Zheng, X. Optoelectronic Crystal of Artificial Atoms in Strain-Textured Molybdenum Disulphide. *Nat. Commun.* **2015**, *6*, 7381.

(14)    Branny, A.; Kumar, S.; Proux, R.; Gerardot, B. D. Deterministic Strain-Induced Arrays of Quantum Emitters in a Two-Dimensional Semiconductor. *Nat. Commun.* **2017**, *8*, 15053.

(15)    Palacios-Berraquero, C.; Kara, D. M.; Montblanch, A. R.-P.; Barbone, M.; Latawiec, P.; Yoon, D.; Ott, A. K.; Loncar, M.; Ferrari, A. C.; Atatüre, M. Large-Scale Quantum-Emitter Arrays in Atomically Thin Semiconductors. *Nat. Commun.* **2017**, *8*, 15093.

(16)    Conley, H. J.; Wang, B.; Ziegler, J. I.; Haglund, R. F.; Pantelides, S. T.; Bolotin, K. I. Bandgap Engineering of Strained Monolayer and Bilayer $MoS_2$. *Nano Lett.* **2013**, *13* , 3626–3630.

(17)    Oakes, L.; Carter, R.; Hanken, T.; Cohn, A. P.; Share, K.; Schmidt, B.; Pint, C. L. Interface Strain in Vertically Stacked Two-Dimensional Heterostructured Carbon-$MoS_2$ Nanosheets Controls Electrochemical Reactivity. *Nat. Commun.* **2016**, *7*, 11796.

(18)    Chari, T.; Ribeiro-Palau, R.; Dean, C. R.; Shepard, K. Resistivity of Rotated Graphite–

Graphene Contacts. *Nano Lett.* **2016**, *16* , 4477–4482.

(19)    Han, Y.; Nguyen, K.; Cao, M.; Cueva, P.; Xie, S.; Tate, M. W.; Purohit, P.; Gruner, S. M.; Park, J.; Muller, D. A. Strain Mapping of Two-Dimensional Heterostructures with Sub-Picometer Precision. *ArXiv180108053 Cond-Mat Physicsphysics* **2018**.

(20)    Lou, S.; Liu, Y.; Yang, F.; Lin, S.; Zhang, R.; Deng, Y.; Wang, M.; Tom, K. B.; Zhou, F.; Ding, H.; Bustillo, K.C.; Wang, X.; Yan, S.; Scott, M.; Minor, A.; Yao, J.; Three-Dimensional Architecture Enabled by Strained Two-Dimensional Material Heterojunction. *Nano Lett.* **2018**, *18* , 1819–1825.

(21)    Kumar, M.; Bhaskaran, H. Ultrasensitive Room-Temperature Piezoresistive Transduction in Graphene-Based Nanoelectromechanical Systems. *Nano Lett.* **2015**, *15* , 2562–2567.

(22)    Barg, A.; Schliesser, A.; Polzik, E. S.; Tsaturyan, Y. Ultracoherent Nanomechanical Resonators via Soft Clamping and Dissipation Dilution. *Nat. Nanotechnol.* **2017**, *12* , 776.

(23)    Lee, J. E.; Ahn, G.; Shim, J.; Lee, Y. S.; Ryu, S. Optical Separation of Mechanical Strain from Charge Doping in Graphene. *Nat. Commun.* **2012**, *3*, 1024.

(24)    Chaste, J.; Saadani, A.; Jaffre, A.; Madouri, A.; Alvarez, J.; Pierucci, D.; Ben Aziza, Z.; Ouerghi, A. Nanostructures in Suspended Mono- and Bilayer Epitaxial Graphene. *Carbon* **2017**, *125* , 162–167.

(25)    Shin, B. G.; Han, G. H.; Yun, S. J.; Oh, H. M.; Bae, J. J.; Song, Y. J.; Park, C.-Y.; Lee, Y. H. Indirect Bandgap Puddles in Monolayer $MoS_2$ by Substrate-Induced Local Strain. *Adv. Mater.* **2016**, *28* , 9378–9384.

(26)    Neumann, C.; Reichardt, S.; Venezuela, P.; Drögeler, M.; Banszerus, L.; Schmitz, M.; Watanabe, K.; Taniguchi, T.; Mauri, F.; Beschoten, B.; Rotkin, S.V.; Stampfer, C. Raman Spectroscopy as Probe of Nanometre-Scale Strain Variations in Graphene. *Nat. Commun.* **2015**, *6*, 8429.

(27)    Yan, W.; He, W.-Y.; Chu, Z.-D.; Liu, M.; Meng, L.; Dou, R.-F.; Zhang, Y.; Liu, Z.; Nie, J.-C.; He, L. Strain and Curvature Induced Evolution of Electronic Band Structures in Twisted Graphene Bilayer. *Nat. Commun.* **2013**, *4*.

(28)    Berciaud, S.; Ryu, S.; Brus, L. E.; Heinz, T. F. Probing the Intrinsic Properties of Exfoliated Graphene: Raman Spectroscopy of Free-Standing Monolayers. *Nano Lett.* **2009**, *9* , 346–352.

(29)    Shivaraman, S.; Barton, R. A.; Yu, X.; Alden, J.; Herman, L.; Chandrashekhar, M.; Park, J.; McEuen, P. L.; Parpia, J. M.; Craighead, H. G.;Spencer, M.G. Free-Standing Epitaxial Graphene. *Nano Lett.* **2009**, *9* , 3100–3105.

(30)    Ni, Z. H.; Chen, W.; Fan, X. F.; Kuo, J. L.; Yu, T.; Wee, A. T. S.; Shen, Z. X. Raman Spectroscopy of Epitaxial Graphene on a SiC Substrate. *Phys. Rev. B* **2008**, *77* , 115416.

(31)    Waldmann, D.; Butz, B.; Bauer, S.; Englert, J. M.; Jobst, J.; Ullmann, K.; Fromm, F.; Ammon, M.; Enzelberger, M.; Hirsch, A.; Maier, S.; Schmuki, P.; Seyller, T.; Spiecker, E.; Weber, H.B. Robust Graphene Membranes in a Silicon Carbide Frame. *ACS Nano* **2013**, *7* , 4441–4448.

(32)    Zabel, J.; Nair, R. R.; Ott, A.; Georgiou, T.; Geim, A. K.; Novoselov, K. S.; Casiraghi, C. Raman Spectroscopy of Graphene and Bilayer under Biaxial Strain: Bubbles and Balloons. *Nano Lett.* **2012**, *12* , 617–621.

(33)    Mohiuddin, T. M. G.; Lombardo, A.; Nair, R. R.; Bonetti, A.; Savini, G.; Jalil, R.; Bonini, N.; Basko, D. M.; Galiotis, C.; Marzari, N.; Novoselov, K.S. ; Geim, A.K. ; Ferrari, A.C. Uniaxial Strain in Graphene by Raman Spectroscopy: G Peak Splitting, Grüneisen Parameters, and Sample Orientation. *Phys. Rev. B* **2009**, *79*, 205433.

(34)    Ni, Z. H.; Yu, T.; Lu, Y. H.; Wang, Y. Y.; Feng, Y. P.; Shen, Z. X. Uniaxial Strain on Graphene: Raman Spectroscopy Study and Band-Gap Opening. *ACS Nano* **2008**, *2* , 2301–2305.

(35)     Chen, S.; Moore, A. L.; Cai, W.; Suk, J. W.; An, J.; Mishra, C.; Amos, C.; Magnuson, C. W.; Kang, J.; Shi, L.; Ruoff, R.S. Raman Measurements of Thermal Transport in Suspended Monolayer Graphene of Variable Sizes in Vacuum and Gaseous Environments. *ACS Nano* **2011**, *5* , 321–328.

(36)     Das, A.; Pisana, S.; Chakraborty, B.; Piscanec, S.; Saha, S. K.; Waghmare, U. V.; Novoselov, K. S.; Krishnamurthy, H. R.; Geim, A. K.; Ferrari, A. C.; Sood, A.K. Monitoring Dopants by Raman Scattering in an Electrochemically Top-Gated Graphene Transistor. *Nat. Nanotechnol.* **2008**, *3* , 210–215.

(37)     Calizo, I.; Balandin, A. A.; Bao, W.; Miao, F.; Lau, C. N. Temperature Dependence of the Raman Spectra of Graphene and Graphene Multilayers. *Nano Lett.* **2007**, *7* , 2645–2649.

(38)     Bruna, M.; Borini, S. Observation of Raman G -Band Splitting in Top-Doped Few-Layer Graphene. *Phys. Rev. B* **2010**, *81* (12).

(39)     Malard, L. M.; Elias, D. C.; Alves, E. S.; Pimenta, M. A. Observation of Distinct Electron-Phonon Couplings in Gated Bilayer Graphene. *Phys. Rev. Lett.* **2008**, *101* .

(40)     Zhao, W.; Tan, P.; Zhang, J.; Liu, J. Charge Transfer and Optical Phonon Mixing in Few-Layer Graphene Chemically Doped with Sulfuric Acid. *Phys. Rev. B* **2010**, *82* .

(41)     Shivaraman, S.; Jobst, J.; Waldmann, D.; Weber, H. B.; Spencer, M. G. Raman Spectroscopy and Electrical Transport Studies of Free-Standing Epitaxial Graphene: Evidence of an AB-Stacked Bilayer. *Phys. Rev. B* **2013**, *87* , 195425.

(42)     Dong, X.; Shi, Y.; Zhao, Y.; Chen, D.; Ye, J.; Yao, Y.; Gao, F.; Ni, Z.; Yu, T.; Shen, Z.; Huang, Y.; Chen, P.; Li, L.-J. Symmetry Breaking of Graphene Monolayers by Molecular Decoration. *Phys. Rev. Lett.* **2009**, *102* , 135501.

(43)     del Corro, E.; Kavan, L.; Kalbac, M.; Frank, O. Strain Assessment in Graphene Through the Raman 2D′ Mode. *J. Phys. Chem. C* **2015**, *119* , 25651–25656.

(44)     Zheng, Q.; Jiang, B.; Liu, S.; Weng, Y.; Lu, L.; Xue, Q.; Zhu, J.; Jiang, Q.; Wang, S.; Peng, L. Self-Retracting Motion of Graphite Microflakes. *Phys. Rev. Lett.* **2008**, *100* , 067205.

(45)     Dienwiebel, M.; Verhoeven, G. S.; Pradeep, N.; Frenken, J. W. M.; Heimberg, J. A.; Zandbergen, H. W. Superlubricity of Graphite. *Phys. Rev. Lett.* **2004**, *92* , 126101.

(46)     Frank, O.; Bousa, M.; Riaz, I.; Jalil, R.; Novoselov, K.S.; Tsoukleri, G.; Parthenios, J.; Kavan, L.; Papagelis, K.; Galiotis, C. Phonon and Structural Changes in Deformed Bernal Stacked Bilayer Graphene. *Nano Lett.,* **2012**, 12, 687–693.

(47)     Gong, L.; Young R.J.; Kinloch, I.A.; Haigh, S.J.; Warner, J.H.; Hinks, J.A.; Xu, Z.; Li L.; Ding F.; Riaz I.; Jalil, R.; Novoselov K.S. Reversible Loss of Bernal Stacking during the Deformation of Few-Layer Graphene in Nanocomposites. *ACS Nano* **2013**, 7, 7287–7294.

(48)     Tsoukleri, G.; Parthenios, J.; Galiotis, C.; Papagelis, K. Embedded trilayer graphene flakes under tensile and compressive loading. *2D Mater.* **2015**, 2, 024009.

(49)     López-Polín, G.; Jaafar, M.; Guinea, F.; Roldán, R.; Gómez-Navarro, C.; Gómez-Herrero, J. The Influence of Strain on the Elastic Constants of Graphene. *Carbon* **2017**, *124*, 42–48.

(50)     Benameur, M. M.; Gargiulo, F.; Manzeli, S.; Autès, G.; Tosun, M.; Yazyev, O. V.; Kis, A. Electromechanical Oscillations in Bilayer Graphene. *Nat. Commun.* **2015**, *6*, 8582.

(51)     Mueller, N.S.; Heeg, S.; Alvarez, M.P.; Kusch, P.; Wasserroth, S.; Clark, N.; Schedin, F.; Parthenios J.; Papagelis.K. Evaluating arbitrary strain configurations and doping in graphene with Raman spectroscopy. *2D Mater.* **2018,** 5, 015016.

**Supporting information**

**Nanomechanical strain concentration on a 2D nanobridge within a large suspended bilayer graphene for molecular mass detection**

Julien Chaste[1*], Amine Missaoui[1], Amina Saadani[1], Daniel Garcia-Sanchez[2], Debora Pierucci[1], Zeineb Ben aziza[1], Abdelkarim Ouerghi[1]

1) Centre de Nanosciences et de Nanotechnologies, CNRS, Univ. Paris-Sud, Universite Paris-Saclay, C2N – Marcoussis
2) Sorbonne Universités, UPMC Univ. Paris 06, CNRS-UMR 7588, Institut des NanoSciences de Paris, F-75005, Paris, France

* julien.chaste@c2n.upsaclay.fr

**S1: Raman difference between doping, strain and temperature dependence**
**S2: Strain concentration 10, 20 and 40μm membranes: additional data**
**S3: Strain simulations**
**S4: Beyond our samples**

## S1: Raman difference between doping, strain and temperature dependence

We have to determine a methodology in order to discriminate Raman peak shift due to the different aspects as strain, doping, temperature, number of layers... We will use the table S1 in order to summary every different aspect.

**Table S1:** Impact of strain, doping or temperature on the Raman G peak shift with also the ratio between the 2D and the G peak, and the G peak splitting behavior [1–9] .

| | $\Delta\omega_{2D}/\Delta\omega_G$ | Splitting peak G | Geometry dependence | Polar dependance | G peak dependence |
|---|---|---|---|---|---|
| Strain | 2.5-3.5 | Yes, for uniaxial strain | yes | yes | 11to32 cm$^{-1}$/% |
| Doping | < 0.7 | Yes, only for n=2.p layers | at the edges | no | 10 cm$^{-1}$/10$^{13}$cm$^{-2}$ |
| Temperature | ~1.7 | No | yes | no | -0.015cm$^{-1}$/K |

## S2: Strain concentration 10, 20 and 40µm membranes: additional datas

To extract the peak position and shift, we have fitted the Raman D and the 2D peak with Voigt function and the G and 2D' peak with a Lorentzian function. Lorentzian lineshape is a usual description of Raman peak. However, the Voigt (or pseudo-Voigt) shape has already been shown to reflect the spectra of graphene better than the Lorentzian, when submitted to a heterogenous broadening by strain or doping or to a Gaussian distribution corresponding to the spectral resolution itself, especially when the sample is suspended [10,11]. For larger membranes, in case of peak splitting, we have fit the G peak with a double Lorentzian call G+ and G-. A peak area is the integration of the signal over the peak shape and the weighted position is the mean frequency position weighted by the Raman intensity of the signal. In the case of G peak shift or position, it means the means value of the G + peak position and the G-peak position.

We can observe the sample to be very clean with the ratio of the G peak over the D peak to be over 40 (Figure S2Ag, S2Bo).

Epitaxial graphene on SiC is well known to have a varying number of layers; it was defined in our case to be between mostly 2 and 3 from the 2D peak shape. A three layers appearance coincides with the SiC step of the substrate before etching. It explains the observation of parallel lines in the Raman data: of the 2D peak intensity, for example, Figure S2Be, S2Cd and of the e-beam images. This aspect has been simulated with COMSOL with a variable thickness membrane but no strong difference was observed in the simulation results for the strain mapping (Figure S2B c and i).

For each measurement, a careful measurement of the spot size was done near the central area (Figure S2B f) over the graphene edge. We have plotted the 2D peak intensity in function of the position along the anvil geometry after the 2D convolution with the laser spot for different diameter. It results in a spot size between 400 and 550nm for all the measurements present here.

Concerning the heating, in Figure 2b (Now 2a), the difference of position for the 2D peak between the 50µW and 100µW is only 0.6cm$^{-1}$. If the 2D peak shift at ~0.025cm$^{-1}$/K, it means we increase the temperature by 24K. If we consider the convolution between the laser spot size and the surface (the heated area is created by the laser itself and of the same order of size) it will increase this value but even if we reach 100K of increase it is not important in graphene. And we only work at 50µW at maximum.

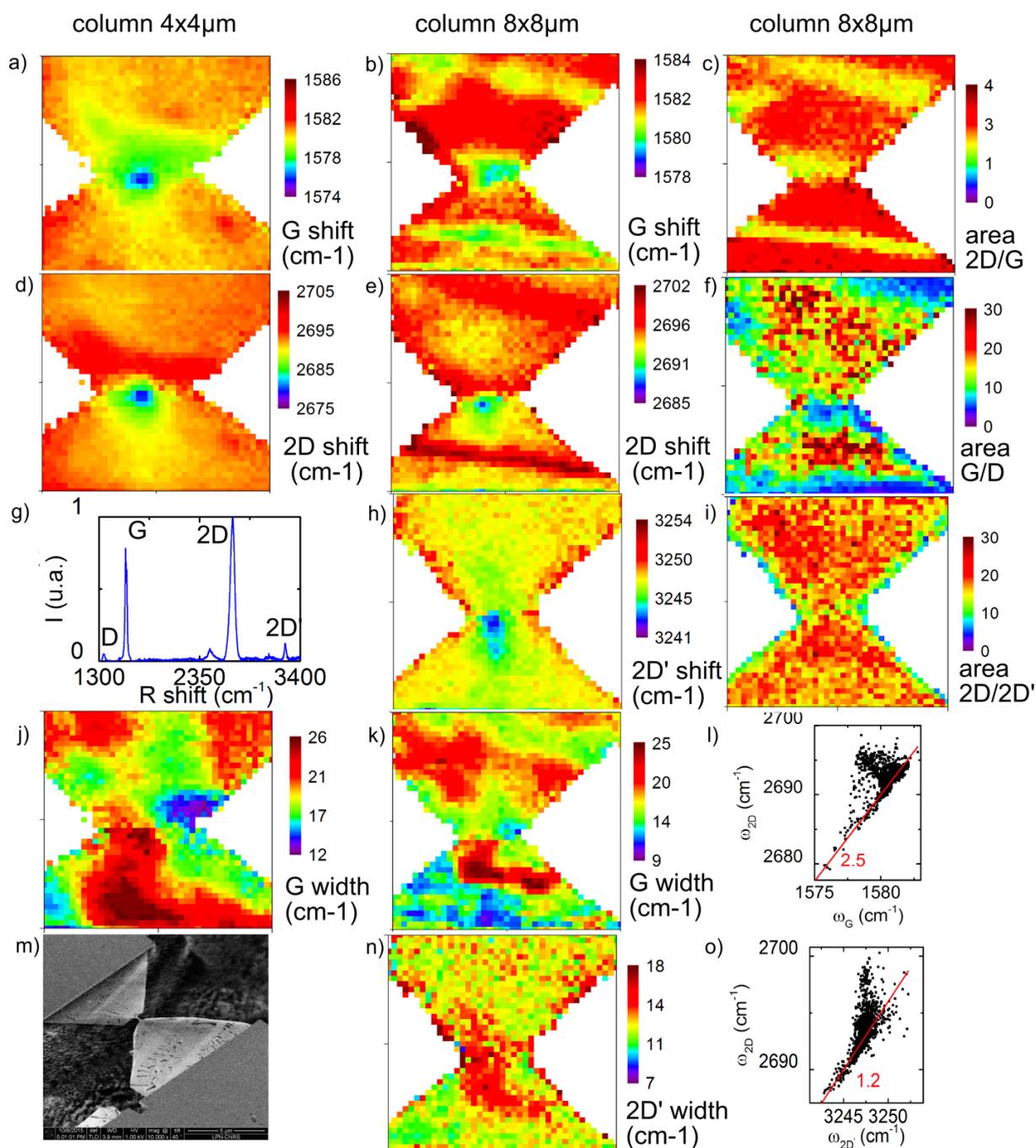

**Figure S2A:** Additional data on the 10µm membrane of Figure 1, **a,d,j)** Raman mapping of 4x4µm. **b,c,e,f,h,i,k,n)** Raman mapping of 8x8µm. **g)** Raman spectrum of the graphene **m)** e-beam image of the membrane. g) Raman spectrum of the different peaks. **l)** Lee et al. diagram of the weighted position of the 2D peak in function of the G peak position with a slope of 2.5 . **o)** Lee et al. diagram of the weighted position of the 2D peak in function of the 2D' peak position with a slope of 1.2.

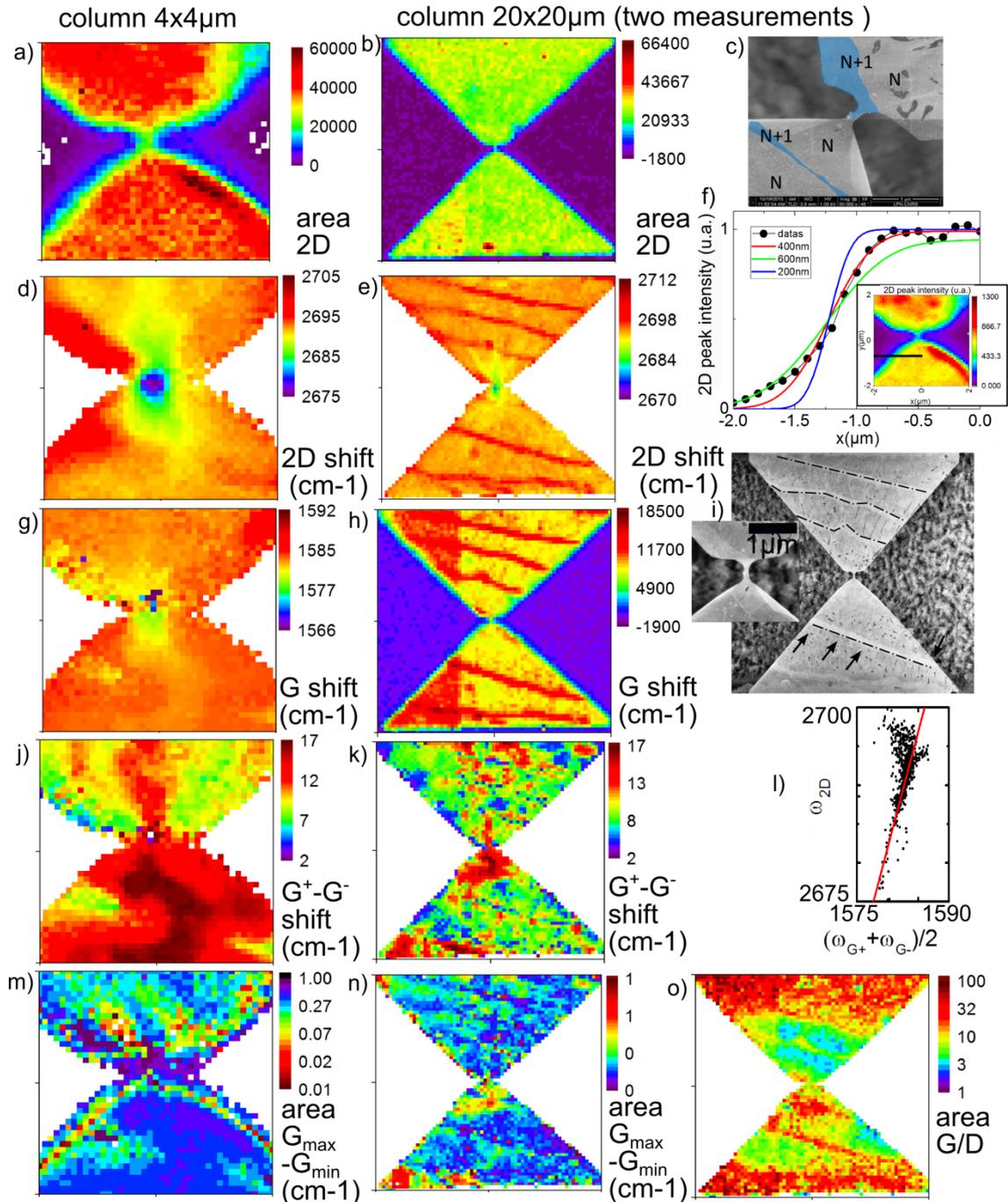

**Figure S2B:** Additional data on the 20μm membrane of Figure 1, **a,d,g,j,m)** Raman mapping of 4x4μm. **b,e,h,k,n,o)** Raman mapping of 20x20μm. **c)** an e-beam image of the sample highlighting the n and n+1 layer's geometries along the membrane. f) Spot size calibration; the 2D peak intensity at the edges of the membrane (black line in the inset) and simulated intensity for different Gaussian spot size. The laser spot size is around 400nm **i)** e-beam image of the membrane **l)** Lee et al. diagram of the weighted position of the 2D peak in function of the G peak position (the eye-guide red line has a slope of 2.2)

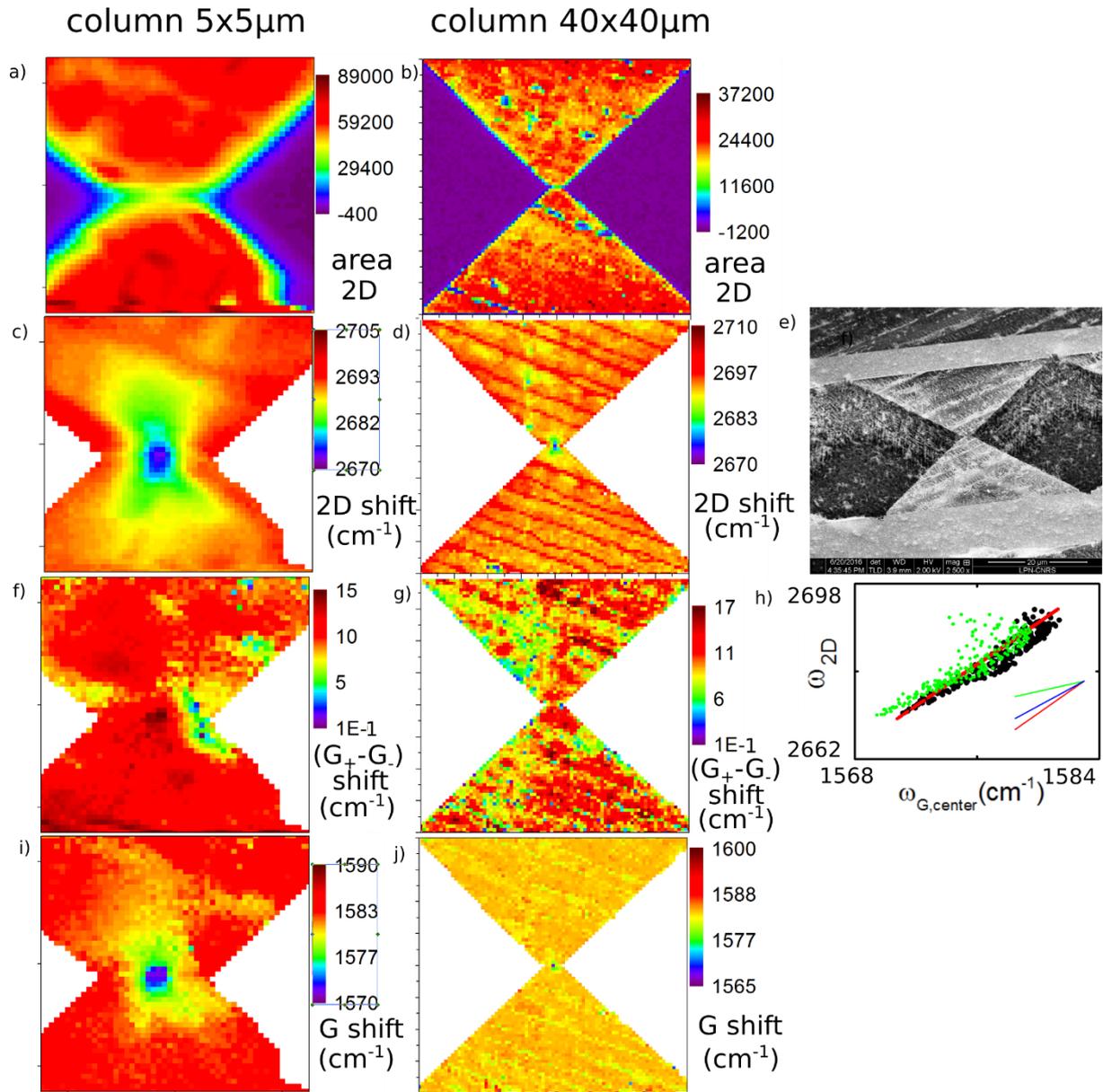

**Figure S2C:** Additional data on the 40μm membrane of Figure 1, **a,c,f,i)** Raman mapping of 5x5μm. **b,d,g,j)** Raman mapping of 40x40μm. **e)** e-beam image of the membrane **o)** Lee et al. diagram of the weighted position of the 2D peak in function of the G peak position (black) and of the 2D maximum intensity position in function of the G maximum intensity position (green).

## S3: Strain simulations

It is not the asymmetry in the Figure 4b that could evidence the concentrated strains at the boundaries because if the strain layer is exactly the opposite signal of the compressed layer, you expect a perfectly symmetric Raman peak but the vertical strain asymmetry between layer is still here and can be very strong. We think we clearly see two parts in the peaks with two peaks maximum (pointed by the two black arrows) in Figure 4b. By asymmetry of the strain, we do not mean and expect an asymmetric Raman response but an asymmetry in the strain itself: one part well above and one part well below the reference peak position at 1580cm⁻¹ (The first one is characteristic of a part under compression and the other under tension).

We don't have a direct evidence of a highly stressed part in our sample but we want to highlight a few thinks about our model;

1) As said before, we measure a Raman response in a broadband window which is around $100cm^{-1}$. This is uncommon and translates a broadening of the Raman spectrum due to strain both in the straining part and in the compressed graphene part.

   If we compare with scenarios taking into account a layer alone (or two symmetric layers with the exact same response), it is impossible to reproduce the broadening and the strain and compressed parts of the spectrum. We tried it a lot to be sure; for example, the Figure S3C in the SI and it is only a small part of our investigation in this direction and even with extreme parameters values, the result does not fit our data very well.

   In fact, a simulation of a strained membrane gives a response mainly below $1580cm^{-1}$ and a simulation of a compressed membrane gives a response mainly above $1580cm^{-1}$, It is an empirical evidence.

2) We have to take into account an asymmetry in the two layers to explain the broadening and the strain versus compression in our data.

3) If we do this, one layer is under compression and the other is under stress.

4) Because the contribution of the stressed part is still nonnegligible below $1470-1480cm^{-1}$ and if we consider our laser convolution, it necessarily implies a part with high strain. And it is the exact opposite for the compressed part.

5) In the end, our scenario of asymmetrical high strain seems to fit well our data.

The concentrated strain at the edges does not affect the spectrum a lot after convolution with the Raman spot size, it seems to be quasi-negligible quantities after convolution with the laser spot size (because it is a local effect and its value is just twice the value in the center part). In fact, it is a good point in our model because this part is obviously the less realistic since our COMSOL simulations do not take into account the z component and possible bending at the edges and at these positions the strain diverge a bit. It is why in our discussion; we talk about the strain in the middle part as the averaged strain.

In order to simulate our data, we have to introduce different parameters as the initial strain for each layer and a frequency shift between the bottom layer and the top layer (certainly due to a doping asymmetry between layers).

We have fixed the geometry of the membrane, the Gruneïsen parameters $\gamma$, the Young modulus E, the Poisson ratio $\nu=0.17$, the layer thickness.

We have been able to determine the shear deformation potential $\beta$, the laser spot size, the peak shape for G and 2D from alternative calibration (as explained below) and we have carefully confirmed these values correspond quite well to the best fitting of our data.

From finite element analyses, we have simulated $\varepsilon_H$ and $\varepsilon_A$ in our membrane we have determined the local Raman peak spectrum before any convolution with the laser distribution. The G peak can split into two peaks under strain G+ and G- with frequency shift define by $\Delta\omega_G=\omega_G.(\gamma. \varepsilon_H \pm 0.5.\beta.\varepsilon_A)$. We have simulated the two peaks by Lorentzian curves with a width of $12cm^{-1}$ and equal amplitudes. This width corresponds quite well to the width of the purple point in Figure 1a, far away from a stress region.

For the 2D peak fit we have used a four peak simulation (A,B,C,D) with $\Delta\omega_{2D}=\omega_{2D}.\gamma_{2D}. \varepsilon_H$. and width $21cm^{-1},28cm^{-1},24cm^{-1},22cm^{-1}$, amplitude of 5.4%, 38.8%, 36.2%,19.6% and frequency of 2650cm⁻¹,2681cm⁻¹,2699cm⁻¹,2715cm⁻¹. $\gamma_{2D}$ =3.(ref [12]). This shape of the 2D curve is typical for the 2D peak in the literature and it is set to fit well the 2D peak far away from the nanobridge in our set-up (without negligible strain).

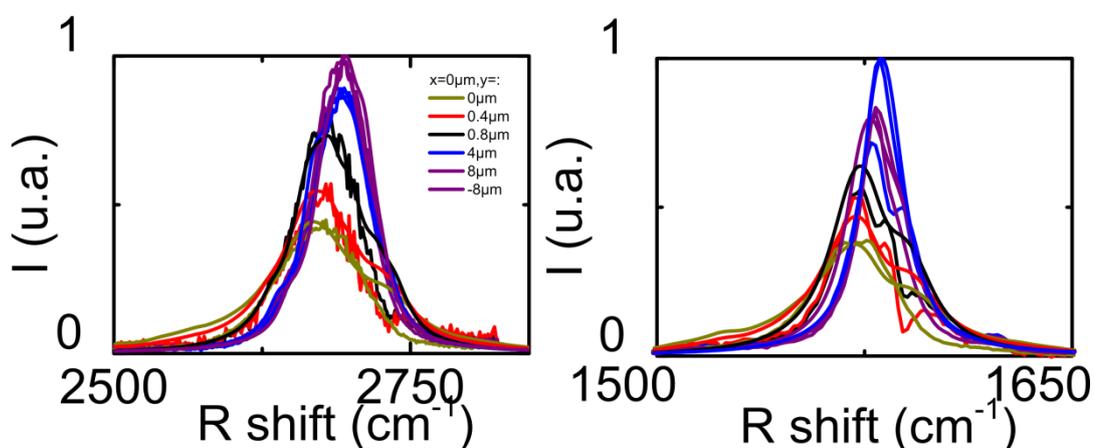

**Figure S3A:** 2D and G peak measurements and simulations on the 40µm membrane for different positions along y. The data are different than in Figure 3; with a larger map (40x40µm) and a lower resolution (laser spot ~750nm). We can reproduce the data quite well with similar parameters than in Figure 3 along with all the membrane.

For the 20µm membrane, the Gaussian spot size was fixed, through simulation, to 500nm and for the 40µm membrane, the Gaussian spot size was fixed to 650nm. It corresponds quite well to the resolution, we measured along the edges of the graphene, Figure S2bf, where intensity must drop from 1 to 0 but is spatially limited by the optical resolution and to a typical resolution of our Raman spectrometer. The difference between 400nm and 650nm is due to the drift of the x,y,z position during the measurements (almost a night per measurement) and was expected from our measurements. The 20µm membrane was measured with a stable position and the 40µm membrane was measured with a small drift during the night.

β is clearly measured to be 0.8. Empirically, it is the dominant parameter which modifies the relation of the maximum intensity position for the Raman G and 2D peaks. It has been used for simulation fitting in addition to the weighted position of the G and 2D peak and the G and 2D peak shape on some specific points.

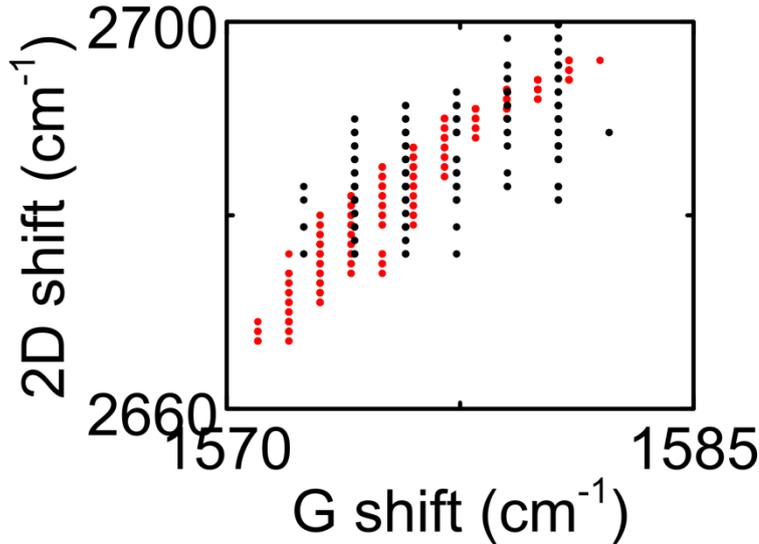

**Figure S3B:** Additional data for the frequency position of the maximum intensity points (FPMIP) around the peak G and 2D. We plot the FPMIP for the 2D peak in function of the FPMIP of the G peak. We have seen empirically the strong dependence of β on the accuracy between the data (black and the simulations (red)) which fix β to 0.8 in our case.

In addition, a ratio of intensities of 71%-29% was empirically introduced in the model between the bottom layer contribution and the upper layer. This ratio is somehow expected for epitaxial graphene[14] and has been introduced in a previous work[14] . It can be due to the doping asymmetry but certainly from diffusive interaction in the bottom layer with the residues of the dangling bonds in epitaxial graphene.

We had also to introduce a frequency shift between the two-layer contributions. We attribute this shifty to a doping asymmetry between layers. For simplification, we have considered this shift to be homogeneous at the scale of the nanobridge, where the strain is applied to the system and where the physic we are concerned here is contained. For the 20µm membrane, we have fixed the non-strained frequency of the G peak to 1582.9 cm$^{-1}$ and 1580.4cm$^{-1}$ for the bottom layer and the top layer respectively. For the 40µm membrane, we have fixed the non-strained frequency of the G peak to 1586.4 cm$^{-1}$ and 1575.6cm$^{-1}$ for the bottom layer and the top layer respectively. For the 2D peak, we have applied the same peak shift. We can observe it corresponds to a bottom layer which is more doped than the top layer as expected from the initial conditions of our epitaxial graphene where the doping of the bottom layer is closed to $2.10^{13}$cm$^{-2}$ before any nanofabrication or etching and the top layer is naturally less doped.

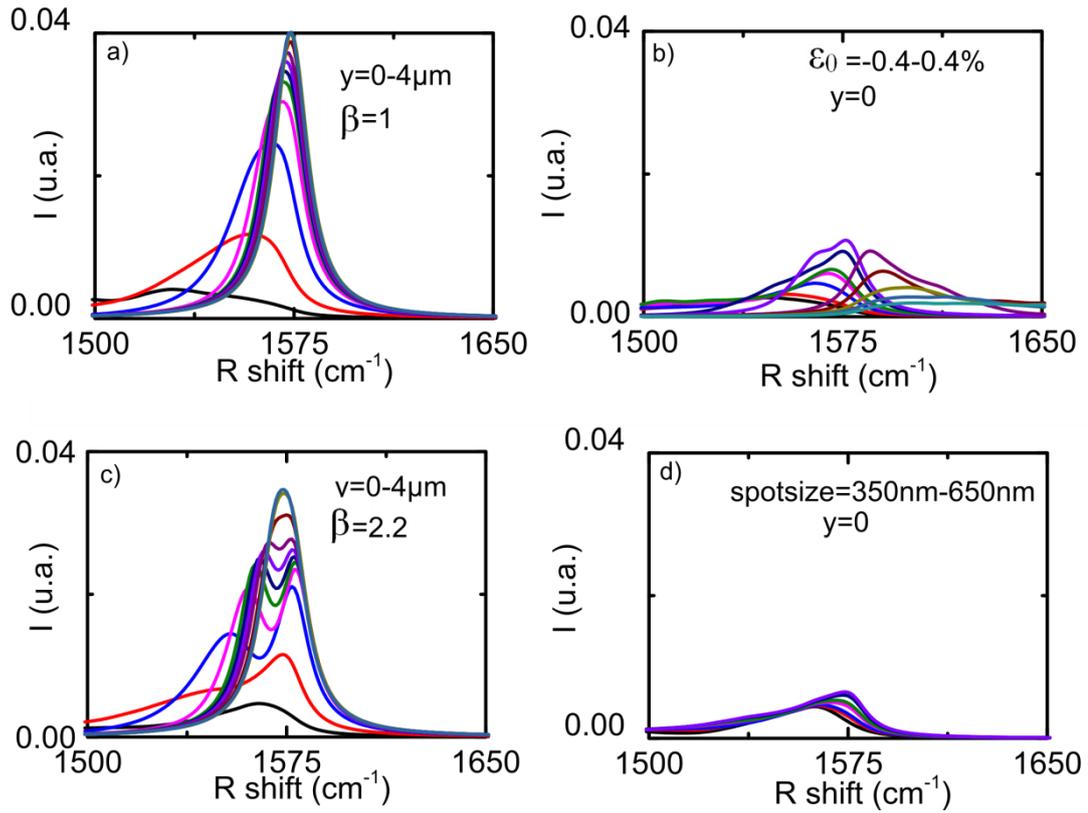

**Figure S3C:** Some examples of simulations for a graphene monolayer with the geometry of $L_0$=20µm for a) different position, b) different initial strain $\varepsilon_0$, a and c) different β d) different spot size. By default spot size=450nm, y=0µm, x=0µm, β=2.2, $\varepsilon_0$=0.15.

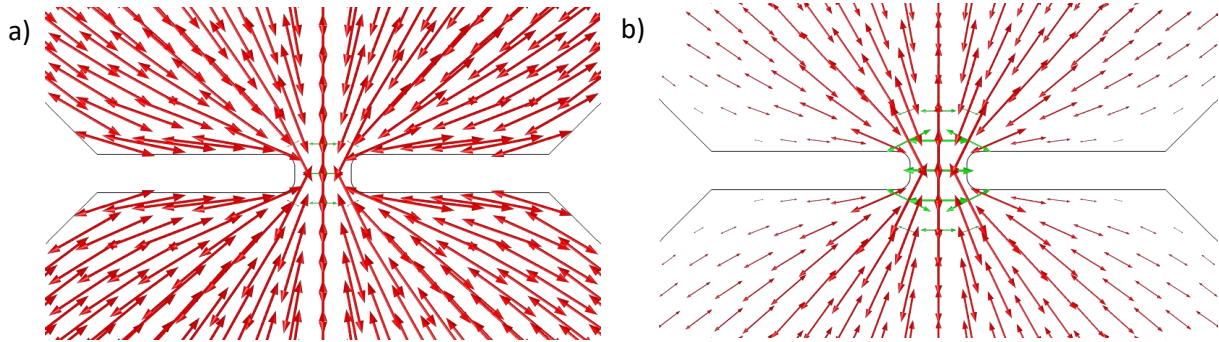

**Figure S3D: Stress tensor simulations of the 40µm membrane a)** We present T1 (in red) and T2 (in green), the eigenvector of the stress tensor along the center of the membrane b) the same plot with the nominal T1 and T2 values (in log scale proportional to 5e-12N/m). This corresponds to the layer under stressed in the main part

## S4: Beyond our samples

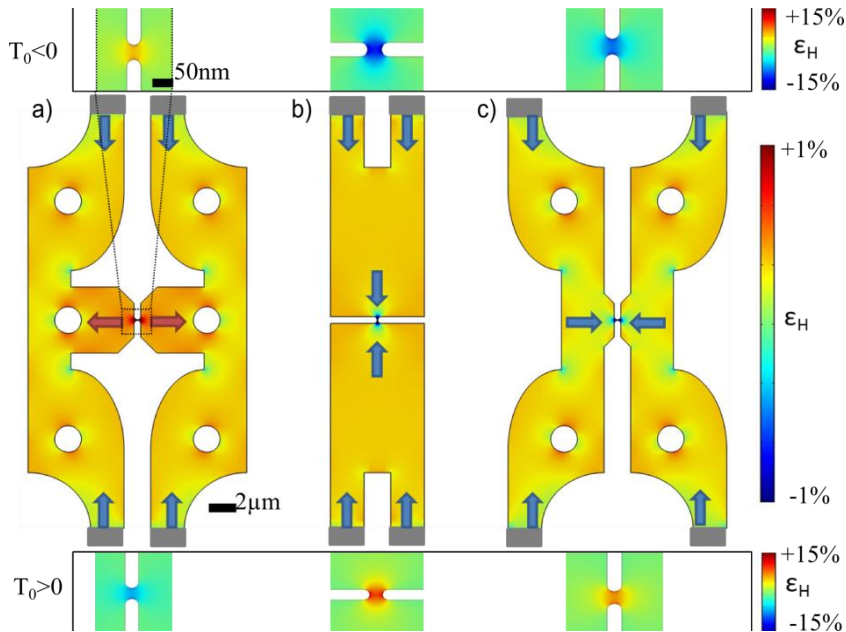

**Figure S4:** Some examples of simulations for a graphene monolayer with 3 different specific geometries. In a) we have a geometry which inverts the strain sign and turn by 90° the strain orientation. In b), we propose a reference geometry with the same length of anchoring edges than in a) and c) without rotation or sign change. In c) we present a geometry with a strain rotation without sign change. Top and bottom parts are zooms on the nanobridge for the initial situation in compression and stressed respectively.

In order to go beyond the anvil cell geometry proposed in the previous samples, we have done additional simulations with monolayer membrane, without buckling strain, with comparable anchoring length and initial strain (+-0.2%). We compare the resulting hydrostatic strain at the center of the nanoconstriction. The value of $\varepsilon_H$ in a) (and c)) are only 50% (75%) of the value the strain at the same position in b). We have these good efficiencies by optimizing the geometry itself. Some holes appeared here in the membrane, in the SI in comparison with the main paper, but it does not strongly affect the result.

## References SOM

(1)     Lee, J. E.; Ahn, G.; Shim, J.; Lee, Y. S.; Ryu, S. Optical Separation of Mechanical Strain from Charge Doping in Graphene. *Nat. Commun.* **2012**, *3*, 1024.

(2)     Das, A.; Pisana, S.; Chakraborty, B.; Piscanec, S.; Saha, S. K.; Waghmare, U. V.; Novoselov, K. S.; Krishnamurthy, H. R.; Geim, A. K.; Ferrari, A. C.; Sood, A.K.  Monitoring Dopants by Raman Scattering in an Electrochemically Top-Gated Graphene Transistor. *Nat. Nanotechnol.* **2008**, *3* , 210–215.

(3)     Bruna, M.; Borini, S. Observation of Raman G -Band Splitting in Top-Doped Few-Layer Graphene. *Phys. Rev. B* **2010**, *81* .

(4)     Ferrari, A. C.; Basko, D. M. Raman Spectroscopy as a Versatile Tool for Studying the Properties of Graphene. *Nat. Nanotechnol.* **2013**, *8* , 235–246.

(5)     Ni, Z. H.; Yu, T.; Lu, Y. H.; Wang, Y. Y.; Feng, Y. P.; Shen, Z. X. Uniaxial Strain on

Graphene: Raman Spectroscopy Study and Band-Gap Opening. *ACS Nano* **2008**, *2* , 2301–2305.

(6)     Mohiuddin, T. M. G.; Lombardo, A.; Nair, R. R.; Bonetti, A.; Savini, G.; Jalil, R.; Bonini, N.; Basko, D. M.; Galiotis, C.; Marzari, N.; Novoselov, K.S.; Geim, A.K.; Ferrari, A.C. Uniaxial Strain in Graphene by Raman Spectroscopy: G Peak Splitting, Grüneisen Parameters, and Sample Orientation. *Phys. Rev. B* **2009**, *79* , 205433.

(7)     Thomsen, C.; Reich, S.; Ordejón, P. *Ab initio* Determination of the Phonon Deformation Potentials of Graphene. *Phys. Rev. B* **2002**, *65* , 073403.

(8)     Calizo, I.; Balandin, A. A.; Bao, W.; Miao, F.; Lau, C. N. Temperature Dependence of the Raman Spectra of Graphene and Graphene Multilayers. *Nano Lett.* **2007**, *7* , 2645–2649.

(9)     Bonini, N.; Lazzeri, M.; Marzari, N.; Mauri, F. Phonon Anharmonicities in Graphite and Graphene. *Phys. Rev. Lett.* **2007**, *99* , 176802.

10) Berciaud, S.; Ryu, S.; Brus, L. E.; Heinz, T. F. Probing the Intrinsic Properties of Exfoliated Graphene: Raman Spectroscopy of Free-Standing Monolayers. *Nano Lett.* *2009*, 9, 346–352.

11) Huang, C.-H.; Lin H.Y.; Huang C.-W.; Liu Y.-M.; Shih F.-Y.; Wang W.-H.; Chui H.-C. Probing substrate influence on graphene by analyzing Raman lineshapes. *Nanoscale Res. Lett.* *2014*, 9, 64.

(12)    Zabel, J.; Nair, R. R.; Ott, A.; Georgiou, T.; Geim, A. K.; Novoselov, K. S.; Casiraghi, C. Raman Spectroscopy of Graphene and Bilayer under Biaxial Strain: Bubbles and Balloons. *Nano Lett.* **2012**, *12* , 617–621.

(13)    Ni, Z. H.; Chen, W.; Fan, X. F.; Kuo, J. L.; Yu, T.; Wee, A. T. S.; Shen, Z. X. Raman Spectroscopy of Epitaxial Graphene on a SiC Substrate. *Phys. Rev. B* **2008**, *77* , 115416.

(14)    Chaste, J.; Saadani, A.; Jaffre, A.; Madouri, A.; Alvarez, J.; Pierucci, D.; Ben Aziza, Z.; Ouerghi, A. Nanostructures in Suspended Mono- and Bilayer Epitaxial Graphene. *Carbon* **2017**, *125*, 162–167.